\documentclass[11pt,a4paper]{article}

\usepackage[T1]{fontenc}
\usepackage[utf8]{inputenc}
\usepackage{lmodern}
\usepackage{amsmath}
\usepackage{amssymb}
\usepackage{graphicx}
\usepackage{booktabs}
\usepackage{array}
\usepackage{siunitx}
\usepackage[numbers]{natbib}
\usepackage[a4paper,margin=1in]{geometry}
\usepackage{hyperref}
\usepackage[protrusion=true,expansion=false]{microtype}
\usepackage{placeins}

\title{INCARBench: A Benchmark for Scientific Configuration\\
in VASP INCAR by Large Language Models}

\author{
\begin{minipage}{0.95\textwidth}
\centering
Bin Shao$^{1,3,4,*}$,
Jixiang Li$^{2}$,
Xinyue Zhang$^{1}$,
Baishun Yang$^{4}$,\\
Zhiyang Liu$^{1,3,4,*}$,
Weichao Wang$^{1,4,6,7,*}$\\[0.5em]
{\small
$^{1}$College of Electronic Information and Optical Engineering, Nankai University, Tianjin 300350, China\\
$^{2}$School of Physics, Nankai University, Tianjin 300071, China\\
$^{3}$Tianjin Key Laboratory of Optoelectronic Sensor and Sensing Network Technology,\\
Nankai University, Tianjin 300350, China\\
$^{4}$Shenzhen Research Institute of Nankai University, Shenzhen 518057, China\\
$^{5}$CIC nanoGUNE BRTA, Tolosa Hiribidea 76, 20018 San Sebasti\'an, Spain\\
$^{6}$College of Energy and Environment Science, Yunnan Normal University,\\
Kunming 650500, China\\
$^{7}$Southwest United Graduate School, Kunming 650092, China\\[0.5em]
$^{*}$Corresponding authors: bshao@nankai.edu.cn; liuzhiyang@nankai.edu.cn;\\
weichaowang@nankai.edu.cn}
\end{minipage}
}

\date{}

\begin{document}
\maketitle

\begin{abstract}
Large language models (LLMs) are increasingly being integrated into first-principles computational workflows, yet their ability to configure scientific calculations remains poorly understood. Here, we introduce INCARBench, a benchmark for evaluating LLMs on input configuration for the \emph{Vienna Ab initio Simulation Package} (VASP) through both configuration generation and repair tasks. Evaluating 19 model configurations reveals substantial capability differences among current frontier models. While several models achieve high semantic and policy accuracy, task-critical correctness remains substantially lower, demonstrating that parameter-level correctness does not necessarily imply scientifically valid configurations. Failure analysis shows that errors concentrate in physically coupled settings involving DFT+$U$, magnetism, and correlated materials, where multiple constraints must be satisfied simultaneously. Repair evaluation further reveals that correcting incorrect settings and preserving already-valid configurations are distinct capabilities, with configuration preservation remaining a major challenge. These findings establish scientific configuration as a measurable capability of large language models and provide a foundation for developing more reliable AI systems for computational materials science.
\end{abstract}


\vspace{1em}

\section{Introduction}

\begin{figure}[htbp]
\centering
\includegraphics[width=0.75\linewidth]{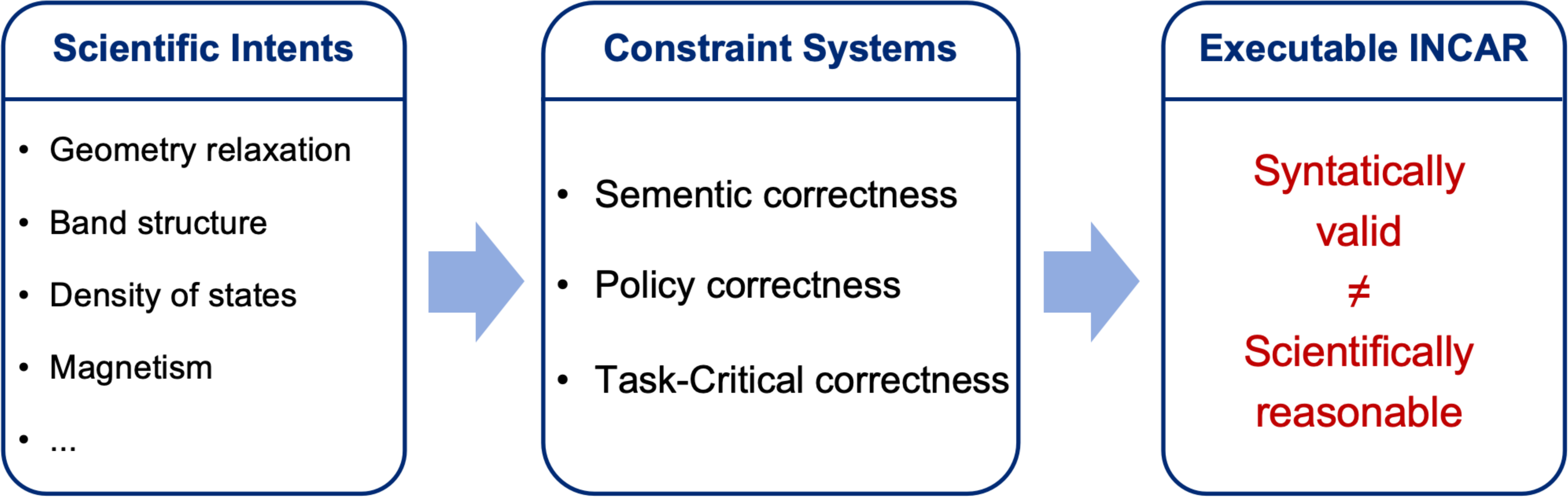}
\caption{Scientific intent encoding in first-principles calculations. Scientific objectives are translated into semantic, policy, and task-critical requirements that jointly determine a VASP INCAR configuration. An executable INCAR file is not necessarily scientifically correct, motivating the evaluation of scientific configuration quality beyond syntax validity and parameter matching.}
\label{fig:intent-encoding}
\end{figure}

Large language models (LLMs) are increasingly being integrated into first-principles computational workflows, assisting with workflow planning, input preparation, execution management, and result interpretation~\cite{lei2024materials,boiko2023autonomous,choi2024accelerating,rubungo2025llmprop,zhang2024honeycomb,guo2025agentic,chen2025vasppilot}. A critical step in these workflows is input configuration: the translation of scientific intent into a computational setup. In the \emph{Vienna Ab initio Simulation Package} (VASP)\cite{kresse1999paw,kresse1996cms,kresse1996prb}, this role is performed by the INCAR file~\cite{vaspwiki}, which specifies the physical model, numerical settings, and workflow behavior of a calculation. Different tasks and physical effects, including geometry optimization, band-structure analysis, magnetism~\cite{hobbs2000noncollinear}, DFT+$U$\cite{dudarev1998lsdau,liechtenstein1995ldaU}, and van der Waals interactions~\cite{grimme2010dftd3,dion2004vdw}, require distinct and often coupled parameter settings. Consequently, an executable INCAR file does not necessarily correspond to a scientifically correct calculation (Figure~\ref{fig:intent-encoding}).

Despite rapid progress in reasoning and scientific applications~\cite{openai2025gpt5,gemini2025,anthropic2025claude,deepseek2025r1,qwen32025,minimax2025m1,kimi2026k25,glm5_2026}, the ability of LLMs to configure first-principles calculations remains largely unexplored. Existing evaluations primarily focus on scientific knowledge, reasoning, property prediction, workflow execution, atomistic modeling capability, or downstream task performance~\cite{zhang2025matscibench,cheung2025msqa,rubungo2024llm4matbench,liu2025mattools,peng2026lambench,rubungo2025llmprop}. They rarely evaluate whether a model can translate scientific intent into a computational configuration that is physically meaningful and workflow consistent. Moreover, practical workflows frequently require modifying existing configurations rather than generating them from scratch, making configuration repair an equally important capability~\cite{guo2025agentic,chen2025vasppilot}.

To address this gap, we introduce INCARBench, a benchmark for evaluating LLMs on VASP INCAR generation and repair. The benchmark assesses whether models can translate scientific intent into computational configurations and modify existing configurations while preserving valid settings. Configuration quality is evaluated through three complementary dimensions: \emph{Semantic Correctness}, \emph{Policy Correctness}, and \emph{Task-Critical Correctness} (TCC), which respectively assess physical intent encoding, numerical-policy selection, and satisfaction of workflow-defining and physics-defining requirements.

\section{Benchmark Design}
\subsection{Benchmark Dataset}

\begin{figure}[htbp]
\centering
\includegraphics[width=0.50\linewidth]{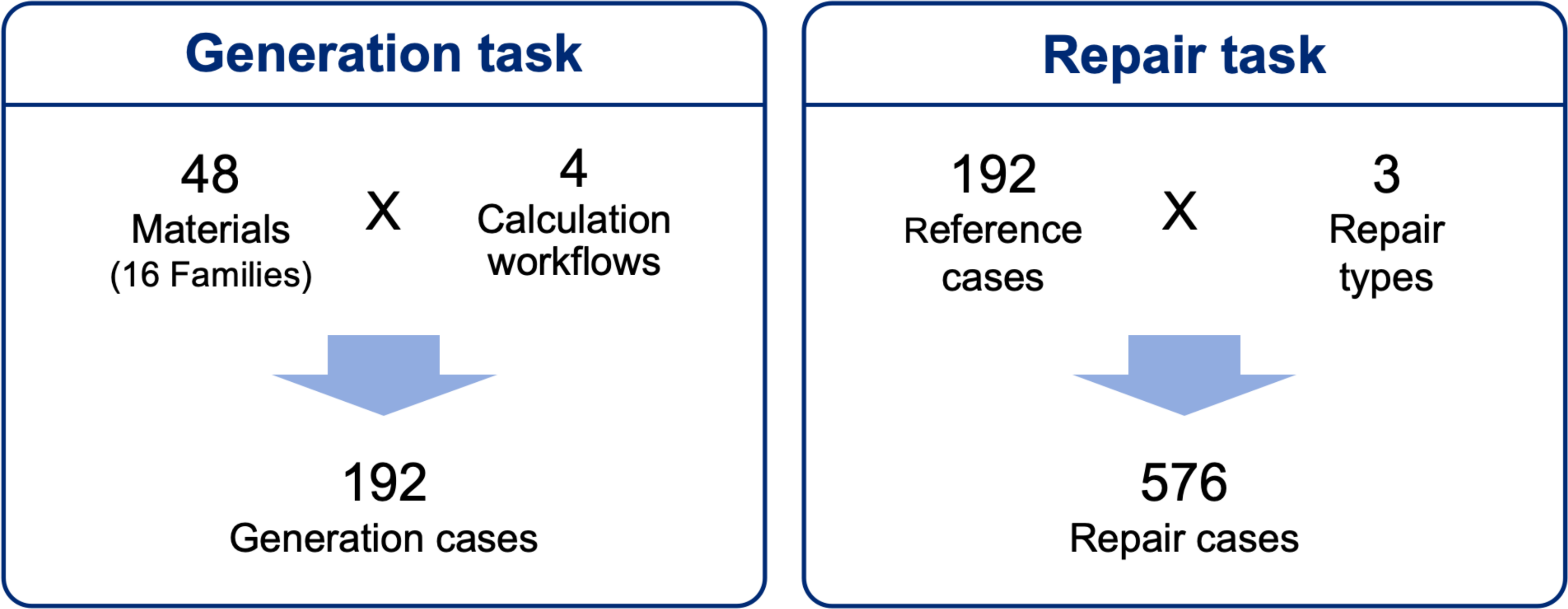}
\caption{Construction of the INCARBench benchmark. The generation task contains 192 cases constructed from 48 materials and four representative workflow types. The repair task is derived directly from the generation cases. Each generation case produces three repair types, resulting in 576 repair cases in total.}
\label{fig:benchmark-design}
\end{figure}

Figure~\ref{fig:benchmark-design} summarizes the design of INCARBench. The generation task contains 48 materials drawn from 16 representative material families and covers a broad range of scientific scenarios commonly encountered in first-principles calculations, including metals, semiconductors, insulators, oxides, correlated materials, layered materials, magnetic materials, and battery materials. The 16 material-family types are listed in the Supplementary Material, Section S1, and the complete benchmark data are available at \url{https://github.com/liuzhiyangnku/incarbench}. Each material is paired with four representative VASP workflow types: static self-consistent field (SCF) calculation, geometry relaxation, band-structure non-self-consistent field (NSCF) calculation, and density-of-states (DOS) NSCF calculation. Applying all four workflow types to all 48 materials yields 192 generation cases.

The repair task is derived directly from the generation task. Each generation case produces three repair variants: a correct-to-preserve variant, a workflow-mistake variant, and a physics-mistake variant. The correct-to-preserve variants contain no injected errors and evaluate a model's ability to avoid over-editing valid configurations, a common failure mode observed in configuration repair. The workflow-mistake variants introduce errors in workflow-defining settings, such as NSCF-related controls and charge-density handling. The physics-mistake variants introduce errors in physics-defining settings, including magnetism, DFT+\textit{U}, van der Waals (vdW) corrections, smearing strategies, and symmetry-sensitive configurations. This results in 576 repair cases in total.

Table~\ref{tab:data-summary} summarizes the benchmark statistics. To support failure localization, each case is assigned one or more physics-aware challenge types. These challenge categories identify scientific settings that require coordinated configuration choices, including NSCF workflows, DFT+\textit{U}, magnetism, van der Waals (vdW) corrections, smearing strategies, and symmetry-sensitive settings. They are used throughout the evaluation to analyze model performance across different scientific regimes and to localize failure modes.

\begin{table}[t]
\centering
\caption{INCARBench dataset summary.}
\label{tab:data-summary}
\small
\begin{tabular}{p{0.30\linewidth}p{0.60\linewidth}}
\toprule
Item & Value \\
\midrule
Materials & 48 materials from 16 material families \\
Workflow types & Static SCF, Geometry Relaxation, Band-Structure NSCF, DOS NSCF \\
Generation cases & 192 \\
Repair cases & 576 \\
Repair variants & Correct-to-Preserve, Workflow-Mistake, Physics-Mistake \\
Challenge types & NSCF Workflow, DFT+U, Magnetism, vdW, Smearing, Symmetry \\
\bottomrule
\end{tabular}
\end{table}

\subsection{Evaluation Framework}

For each benchmark case, the reference INCAR is derived from Materials Project VASP calculation records~\cite{jain2013materialsproject,horton2025materialsproject,ong2013pymatgen} and normalized for benchmark evaluation. Because multiple INCAR configurations may correspond to scientifically reasonable calculations, the reference is treated as a standardized evaluation target rather than a unique ground truth. Detailed benchmark construction, normalization rules, parameter classifications, and benchmark definitions are provided in the Supplementary Material, Sections S1--S3.

Generated INCAR files are compared against the normalized reference configuration. INCARBench evaluates configuration quality through three complementary dimensions: semantic correctness, policy correctness, and TCC. Semantic correctness measures whether the intended physical meaning of a calculation is correctly encoded. Policy correctness evaluates whether numerical and methodological choices are appropriately configured. TCC evaluates whether all task-critical requirements of a calculation are simultaneously satisfied. Together, these dimensions distinguish parameter-level correctness from scientifically meaningful configuration quality.

Parameters are grouped into three categories: semantic parameters, policy parameters, and optional parameters. Task-critical parameter groups are defined separately and consist of semantic and policy parameters whose incorrect configuration would invalidate the intended calculation. When omitted parameters have explicit default values defined in the VASP documentation~\cite{vaspwiki}, those defaults are applied during normalization and comparison. Semantic parameters determine the physical meaning of a calculation and include settings associated with physical effects and workflow intent. Representative examples include \texttt{ISPIN}, \texttt{MAGMOM}, \texttt{LDAU}, and \texttt{ICHARG}. Policy parameters define numerical and methodological choices, including \texttt{ENCUT}, \texttt{EDIFF}, \texttt{SIGMA}, and \texttt{NELM}. Optional parameters are retained for completeness and reproducibility but are not considered task-defining requirements. Examples include auxiliary control tags such as \texttt{ADDGRID} and \texttt{LREAL}.

The benchmark first evaluates parameter-level correctness. Let $P_{\mathrm{sem}}$ and $P_{\mathrm{pol}}$ denote the sets of semantic and policy parameters, respectively. A semantic parameter is counted as correct only when the normalized prediction exactly matches the reference value,

\begin{equation}
S_{\mathrm{sem}}
=
\frac{1}{|P_{\mathrm{sem}}|}
\sum_{p\in P_{\mathrm{sem}}}
\mathbf{1}\left[\hat{p}=p^\ast\right],
\end{equation}
where $\hat{p}$ and $p^\ast$ denote the predicted and reference values, respectively. For policy parameters, correctness is evaluated using parameter-specific equivalence or tolerance rules,

\begin{equation}
S_{\mathrm{pol}}
=
\frac{1}{|P_{\mathrm{pol}}|}
\sum_{p\in P_{\mathrm{pol}}}
\mathbf{1}\left[d_p(\hat{p},p^\ast)\le \tau_p\right],
\end{equation}
where $d_p(\hat{p},p^\ast)$ denotes the comparison rule associated with parameter $p$ and $\tau_p$ is the corresponding tolerance threshold. In these equations, $\mathbf{1}[\cdot]$ denotes the indicator function, which equals 1 when the condition inside the brackets is satisfied and 0 otherwise. Semantic and policy scores are reported separately because different models often exhibit distinct strengths in physical reasoning and numerical configuration. The overall parameter-level score is defined as

\begin{equation}
S_{\mathrm{overall}}
=
\frac{S_{\mathrm{sem}}+S_{\mathrm{pol}}}{2}.
\end{equation}

INCARBench additionally reports a TCC metric. TCC evaluates whether all task-critical parameter groups are simultaneously correct. These groups correspond to workflow-defining and physics-defining requirements whose incorrect configuration would invalidate the intended calculation despite otherwise high parameter-level accuracy. Representative examples include correlated-electron treatment (\texttt{LDAU}, \texttt{LDAUL}, \texttt{LDAUU}), magnetic initialization (\texttt{ISPIN}, \texttt{MAGMOM}), van der Waals corrections (\texttt{IVDW}), occupation treatment (\texttt{ISMEAR}), and workflow-dependent charge-density controls (\texttt{ICHARG}). Unlike semantic and policy scores, TCC is a strict case-level metric: a single error in any task-critical parameter group causes the entire case to fail.

\subsection{Evaluation Protocol}

All models are evaluated using identical task definitions, prompt templates, parsing procedures, and scoring scripts. Model outputs are converted into structured key--value dictionaries before evaluation. Parameter classifications, normalization rules, comparison criteria, and evaluation targets are fixed prior to benchmarking and remain unchanged for all models. This protocol ensures consistent and reproducible comparisons across generation and repair tasks.

A total of 19 frontier LLM configurations were evaluated, including models from OpenAI, Anthropic, Google, DeepSeek, Alibaba, Moonshot AI, MiniMax, Zhipu AI, and Meta. For models supporting configurable reasoning modes, multiple variants were evaluated. Throughout the paper, abbreviated display names are used in figures for readability (see Table~\ref{tab:setup}). Specifically, GPT-5.4 XH denotes the extended-reasoning configuration and GPT-5.4 NR denotes the no-reasoning configuration. Similarly, Qwen3.6+ T and Qwen3.6+ NT denote the thinking and no-thinking variants, respectively, whereas Kimi K2.5 T and Kimi K2.5 NT denote the corresponding reasoning-enabled and reasoning-disabled variants.

\begin{table}[t]
\centering
\caption{Model names used throughout the paper.}
\label{tab:setup}
\begin{tabular}{p{0.24\linewidth}p{0.66\linewidth}}
\hline
Display Name & Model Configuration \\
\hline
GPT-5.4 & GPT-5.4 (Default)~\cite{openai2025gpt5} \\
GPT-5.4 XH & GPT-5.4 (Extended Reasoning)~\cite{openai2025gpt5} \\
GPT-5.4 NR & GPT-5.4 (No Reasoning)~\cite{openai2025gpt5} \\
GPT-5.3 & GPT-5.3 (Codex)~\cite{openai2025gpt5} \\
Claude Opus & Claude Opus 4.6~\cite{anthropic2026claude46} \\
Claude Sonnet & Claude Sonnet 4.6~\cite{anthropic2026claude46} \\
Claude Haiku & Claude Haiku 4.5~\cite{anthropic2025haiku45} \\
Gemini 3.1 & Gemini 3.1 Pro Preview~\cite{gemini2026gemini31} \\
Gemini 2.5 & Gemini 2.5 Pro~\cite{gemini2025} \\
DeepSeek R1 & DeepSeek-R1~\cite{deepseek2025r1} \\
DeepSeek V3 & DeepSeek-V3~\cite{deepseek2024v3} \\
Qwen3.6+ T & Qwen3.6+ (Thinking)~\cite{qwen32025} \\
Qwen3.6+ NT & Qwen3.6+ (No-Thinking)~\cite{qwen32025} \\
Qwen Coder & Qwen3-Coder-Plus~\cite{qwen2025coder} \\
GLM-5.1 & GLM-5.1~\cite{glm5_2026} \\
Kimi K2.5 T & Kimi K2.5 (Thinking)~\cite{kimi2026k25} \\
Kimi K2.5 NT & Kimi K2.5 (No-Thinking)~\cite{kimi2026k25} \\
MiniMax & MiniMax-M2.7~\cite{minimax2026m27} \\
Llama 4 & Llama 4 Maverick~\cite{meta2025llama4} \\
\hline
\end{tabular}
\end{table}

\section{Results}

\subsection{Benchmark Performance}

\begin{figure}[htbp]
\centering
\includegraphics[width=\linewidth]{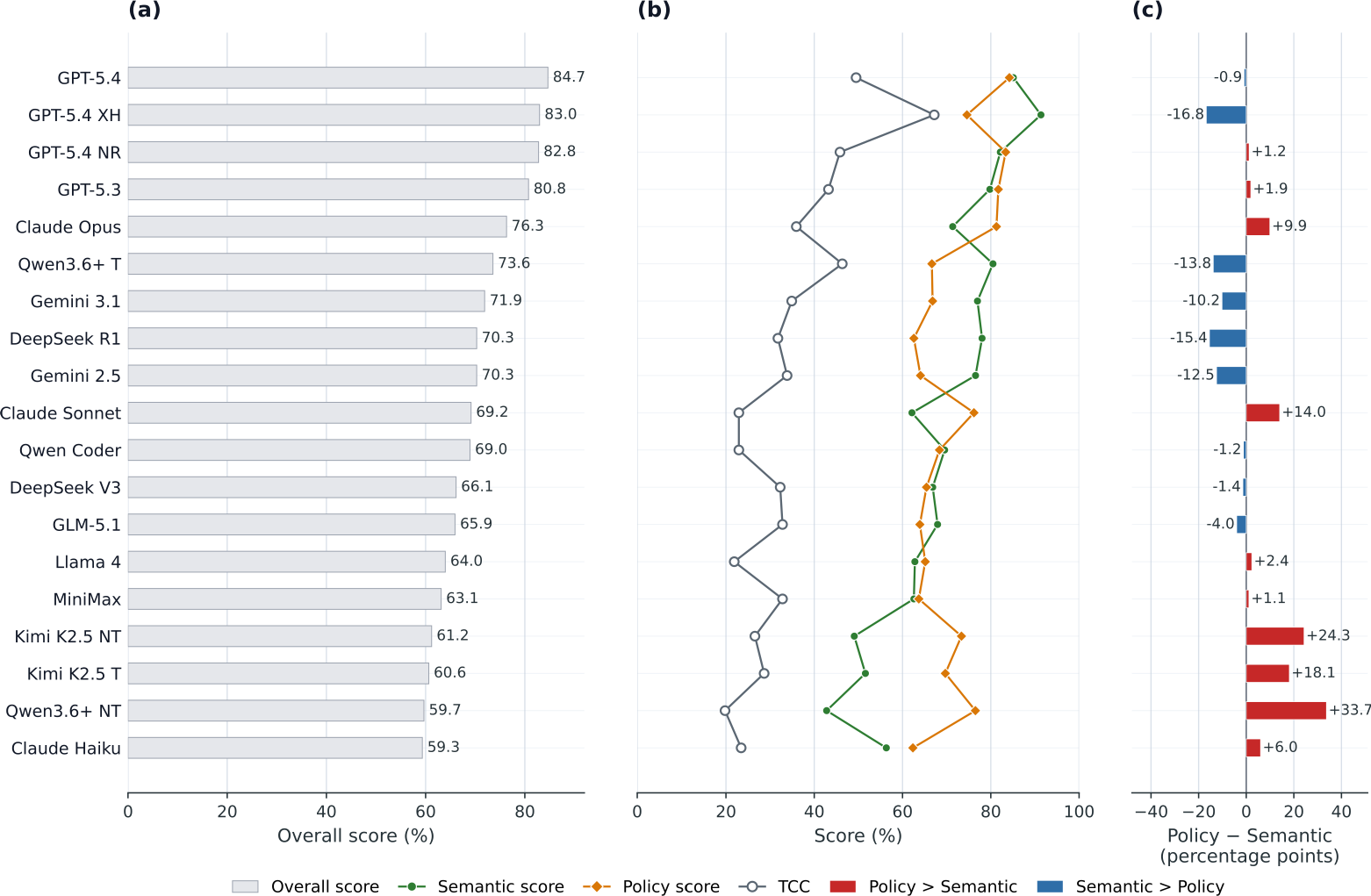}
\caption{
Generation performance on INCARBench. (a) Overall generation score. The broad score distribution demonstrates substantial capability separation across models, with GPT-5.4 achieving the highest overall performance. (b) Semantic correctness, policy correctness, and task-critical correctness (TCC). TCC remains consistently lower than semantic and policy correctness, indicating that parameter-level accuracy does not necessarily imply task-critical correctness. (c) Policy$-$Semantic gap, defined as the difference between policy and semantic correctness. Positive values indicate stronger policy fidelity than semantic understanding, whereas negative values indicate stronger semantic understanding than policy fidelity. The wide distribution of gaps reveals that physical understanding and numerical-policy recovery represent partially independent capabilities in scientific configuration tasks.
}
\label{fig:generation-ranking}
\end{figure}

Figure~\ref{fig:generation-ranking}(a) shows clear capability separation across models. GPT-5.4 achieves the highest overall score (84.7\%), followed by GPT-5.4 XH and GPT-5.4 NR. The broad score distribution, spanning more than 25 percentage points between the strongest and weakest models, indicates that INCARBench is far from saturation and can effectively discriminate scientific configuration capabilities. Achieving a high score requires not only selecting appropriate parameter values but also consistently coordinating physical assumptions, numerical strategies, and workflow-specific constraints. Additional details on workflow types, material families, challenge types, and scoring rules are provided in the Supplementary Material, Sections S1--S3.

Figure~\ref{fig:generation-ranking}(b) further decomposes performance into semantic correctness, policy correctness, and task-critical correctness (TCC). Although leading models achieve strong semantic and policy performance, TCC remains substantially lower across nearly all model families. This result demonstrates that parameter-level correctness does not necessarily translate into task-critical correctness. A model may correctly configure most individual parameters while still failing a small number of workflow-defining or physics-defining parameter groups. Because TCC requires all task-critical settings to be simultaneously correct, even a limited number of mistakes can invalidate an otherwise high-scoring configuration. The persistent TCC deficit therefore highlights the distinction between parameter-level accuracy and scientifically meaningful configuration quality.

Figure~\ref{fig:generation-ranking}(c) reveals a second pattern: semantic correctness and policy correctness are only partially coupled. The Policy$-$Semantic gap varies substantially across models, ranging from strongly negative values for GPT-5.4 XH, DeepSeek R1, and Gemini 2.5 to strongly positive values for Qwen3.6+ NT, Kimi K2.5 NT, and Kimi K2.5 T. Models with positive gaps achieve stronger policy performance than semantic performance, suggesting that they reproduce common parameter patterns more successfully than they capture the underlying physical intent of the calculation. In contrast, models with negative gaps exhibit stronger semantic understanding than policy fidelity, indicating that they often identify the correct physical workflow while failing to fully recover the corresponding numerical strategy. This dissociation reveals two distinct failure modes and suggests that scientific configuration requires both physical understanding and policy fidelity rather than either capability alone.

\FloatBarrier

\begin{figure}[htbp]
\centering
\includegraphics[width=\linewidth]{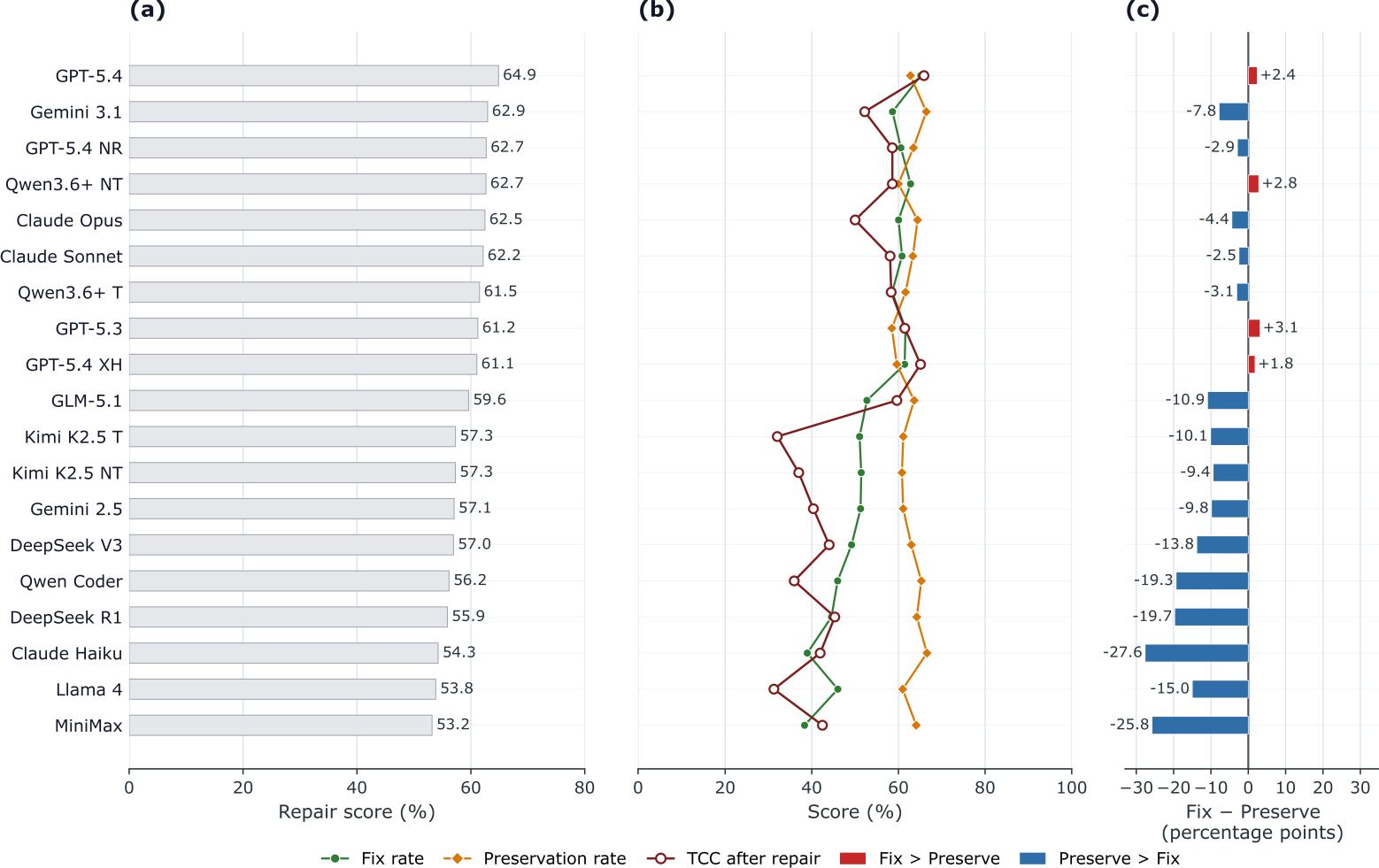}
\caption{
Repair performance on INCARBench. (a) Overall repair score ranked by aggregate repair performance. GPT-5.4 achieves the highest overall repair score, although substantial separation remains across models. (b) Capability decomposition into error-fix rate, preservation rate, and task-critical correctness (TCC) after repair. Preservation rates are generally higher than error-fix rates, while TCC remains substantially lower than both component-level metrics, indicating that successful local edits do not necessarily restore a task-critical correct configuration. (c) Fix$-$Preserve gap. Positive values indicate aggressive repair behavior, whereas negative values indicate conservative repair behavior. The predominance of negative gaps shows that preservation generally exceeds error correction, highlighting configuration preservation as a major bottleneck in scientific repair.
}
\label{fig:repair-ranking}
\end{figure}

Repair performance remains far from saturation across all evaluated models (Figure~\ref{fig:repair-ranking}). As shown in Figure~\ref{fig:repair-ranking}(a), substantial separation exists between the strongest and weakest repair systems, indicating that reliable INCAR repair remains a challenging task even for frontier models. GPT-5.4 achieves the highest overall repair score, followed by Gemini 3.1 and GPT-5.4 NR. Unlike generation, repair requires models not only to identify incorrect settings but also to avoid unnecessary modifications to already-correct configurations.

Figure~\ref{fig:repair-ranking}(b) decomposes repair performance into error-fix rate, preservation rate, and TCC after repair. Across most models, preservation rates are consistently higher than error-fix rates, indicating that maintaining valid configurations is generally easier than correcting injected errors. However, TCC remains substantially lower than both component-level repair metrics. Many models successfully preserve large portions of the original configuration and repair a substantial fraction of injected errors, yet still fail to restore all task-critical parameter groups simultaneously. This result highlights the difference between local parameter correction and recovery of a scientifically valid configuration. Repair construction and repair metrics are described in the Supplementary Material, Section S4.

A third observation is the existence of distinct repair behaviors. Figure~\ref{fig:repair-ranking}(c) reports the Fix$-$Preserve gap, which measures the balance between active error correction and configuration preservation. Most models exhibit negative values, indicating that preservation generally exceeds error correction. Large negative gaps suggest conservative repair behavior, where models protect existing configurations but leave many injected errors unresolved. In contrast, a small number of models exhibit near-zero or slightly positive gaps, reflecting a more aggressive editing strategy that improves correction performance while increasing the risk of modifying valid settings. Together, these results show that error correction and configuration preservation are distinct capabilities. Preservation remains a major bottleneck in scientific repair, and effective repair requires both accurate diagnosis of incorrect settings and careful restraint against unnecessary modifications.

\FloatBarrier

\subsection{Failure Localization}

Overall benchmark performance reveals substantial differences across models, but it does not explain where failures occur. To localize failure modes, we aggregate results across all evaluated models and analyze performance jointly across material families and challenge types. Rather than asking which model performs best, this analysis identifies the scientific regimes that remain difficult for current LLMs.

\begin{figure}[htbp]
\centering
\includegraphics[width=0.75\linewidth]{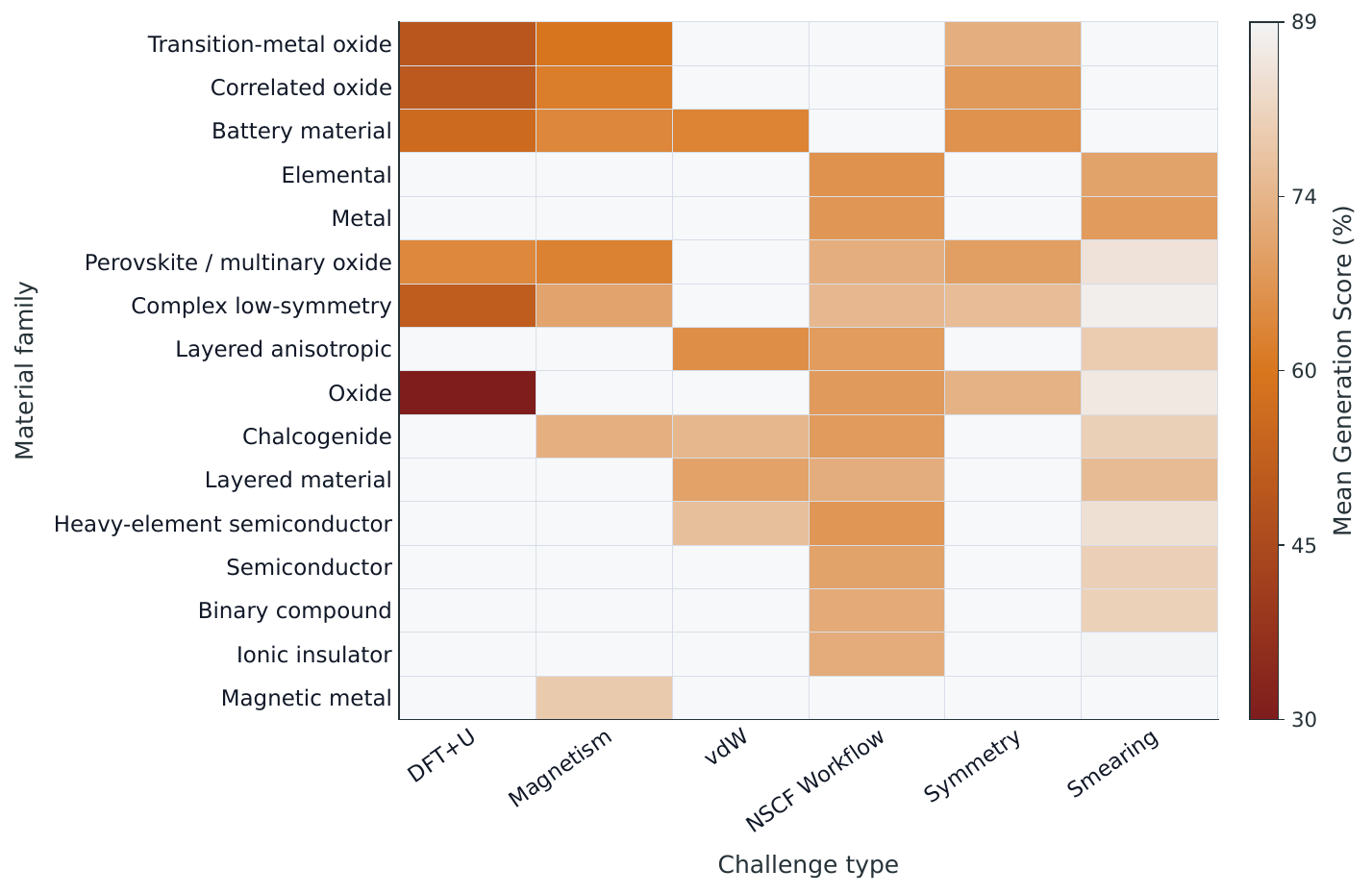}
\caption{
Material--challenge failure landscape. Cell colors indicate the mean generation score averaged across all evaluated models. Darker regions correspond to more difficult scientific regimes. Failures are concentrated in specific material--challenge combinations rather than being uniformly distributed across the benchmark. Transition-metal oxides, correlated oxides, and battery materials remain particularly challenging in DFT+\textit{U}-related settings, while magnetic and layered materials exhibit lower performance under magnetism- and vdW-related constraints, respectively. The results show that benchmark difficulty arises primarily from interactions between material-dependent physics and challenge-specific requirements.
}
\label{fig:failure-landscape}
\end{figure}

Figure~\ref{fig:failure-landscape} reveals a highly structured failure landscape. Difficulties are concentrated in specific material--challenge combinations rather than being uniformly distributed across the benchmark. The most challenging regions occur in transition-metal oxides, correlated oxides, and battery materials when DFT+\textit{U}-related constraints are present. Magnetic materials become substantially more difficult under magnetic configurations, while layered materials exhibit reduced performance in vdW-related settings. In contrast, semiconductors, binary compounds, and ionic insulators generally remain easier across most challenge categories.

A second observation is that challenge difficulty is strongly context dependent. For example, DFT+\textit{U} is not uniformly difficult across all materials but becomes particularly challenging when combined with correlated and transition-metal oxide systems. Similarly, vdW-related difficulties are concentrated in layered materials rather than appearing broadly across the benchmark. These patterns indicate that failures arise primarily from interactions between material-dependent physics and challenge-specific constraints rather than from isolated parameter prediction errors.

Overall, the failure landscape suggests that current LLM limitations emerge most strongly in physically coupled scientific regimes where semantic correctness, policy correctness, and task-critical correctness must be satisfied simultaneously. This result helps explain why parameter-level accuracy often remains substantially higher than task-critical correctness. Challenge-type definitions are provided in the Supplementary Material, Section S1, and representative case-level failure examples are shown in Supplementary Figure S1 and discussed in Supplementary Section S5.

\FloatBarrier

\section{Discussion}

INCARBench treats first-principles input configuration as a scientific capability rather than a formatting task. The results show that generating an executable INCAR file is substantially easier than producing a scientifically valid configuration. This distinction is important because many workflow-defining and physics-defining requirements emerge from interactions among multiple parameters rather than from individual tags in isolation.

More broadly, the benchmark suggests that scientific configuration quality cannot be adequately characterized by parameter-level accuracy alone. The observed gaps among semantic correctness, policy correctness, and task-critical correctness indicate that physical understanding, numerical-policy selection, and workflow consistency represent related but distinct capabilities. As a result, conventional parameter-matching metrics may overestimate the ability of LLMs to support real scientific workflows.

INCARBench currently focuses on INCAR configuration and does not evaluate execution outcomes, convergence behavior, or calculated properties. Future work may extend this framework toward execution-aware and workflow-level benchmarks that more directly connect configuration quality with scientific outcomes.

\section{Conclusion}

We introduced INCARBench, a benchmark for evaluating large language models on VASP INCAR generation and repair. The benchmark reveals persistent gaps between parameter-level correctness and scientifically valid configurations, particularly in physically coupled settings such as DFT+\textit{U}, magnetism, and correlated materials. It also shows that error correction and configuration preservation are distinct capabilities in scientific repair. These findings establish scientific configuration as a measurable capability of large language models and provide a foundation for more reliable AI-driven scientific workflows. All benchmark data have been uploaded to GitHub at \url{https://github.com/liuzhiyangnku/incarbench}.

\section*{Acknowledgements}

This work was supported by the Yunan Provincial Science and Techonology United Graduate School (202402AO370001), the National Natural Science Foundation of China (Grant No. 12474235 and 22576107), the Key Program of Tianjin Natural Science Foundation (Grant No. 24JCZDJC01230), and the Supercomputing Center of Nankai University (NKSC).

\bibliographystyle{unsrtnat}
\bibliography{references}

\end{document}